\documentclass[letter]{aa}
\usepackage{graphicx}
\usepackage{natbib}
\begin{document}

\title{Bayesian analysis of the radial velocities of HD 11506 reveals another planetary companion}

\author{Mikko Tuomi\inst{1,}\inst{2}\thanks{\email{mikko.tuomi@utu.fi}} \and Samuli Kotiranta\inst{2}\thanks{\email{samuli.kotiranta@utu.fi}}}

\institute{Department of Mathematics and Statistics, P.O.Box 68, FI-00014 University of Helsinki, Finland \and Tuorla Observatory, Department of Physics and Astronomy, University of Turku, FI-21500, Piikki\"o, Finland}

\date{Received xx.xx.xxxx / Accepted xx.xx.xxxx}

\abstract
{}
{We aim to demonstrate the efficiency of a Bayesian approach in
analysing radial velocity data by reanalysing a set of radial velocity measurements.}
{We present Bayesian analysis of a recently published set of radial
velocity measurements known to contain the signal of one extrasolar
planetary candidate, namely, HD 11506. The analysis is conducted using
the Markov chain Monte Carlo method and the resulting distributions of
orbital parameters are tested by performing direct integration of
randomly selected samples with the Bulirsch-Stoer method. The
magnitude of the stellar radial velocity variability, known as jitter,
is treated as a free parameter with no assumptions about its magnitude.}
{We show that the orbital parameters of the planet known to be present
in the data correspond to a different solution when the jitter is
allowed to be a free parameter. We also show evidence of an additional
candidate, a 0.8~$M_{Jup}$ planet with period of about 0.5~yr in orbit
around HD 11506. This second planet is inferred to be present with a
high level of confidence.}
{}

\keywords{planetary systems -- Methods: statistical -- Techniques: radial velocities -- Stars: individual: HD 11506}

\titlerunning{Bayesian analysis of the radial velocities of HD 11506}

\authorrunning{M. Tuomi and S. Kotiranta}

\maketitle

\section{Introduction}

The question of whether an extrasolar planet is detectable or not
depends on a delicate mixture of observational technology, effective
tools for data analysis, and theoretical understanding on the related
phenomena. Traditionally, the instrumentation has been the most
celebrated part of this trinity \citep{santos,li}, but the fainter the
signals that one is able to detect, the more the attention should be
directed towards the two other factors involved in successful
discoveries. It is already known that of poor understanding of the
noise-generating physics may produce misleading results
\citep{queloz}, although this is also true for the means the data
analysis. In many cases the statistical methods involved in the data
reduction and analysis of the observations have a tendency to identify
local solutions instead of global ones. This is particularly true
especially when information is extracted using gradient-based algorithms.

In this letter, we reanalyze the radial velocity (RV) data of a
detected extrasolar planet host, namely HD 11506 \citep{fischer}. Our
analysis is based on Bayesian model probabilities and full inverse solutions, i.e., the full \emph{a posteriori} probability densities of model parameters. The Bayesian model probabilities provide a strict mathematical criterion for deciding how many planetary signals are present in the data. We calculate the inverse solutions using the Markov chain Monte Carlo (MCMC) method with the Metropolis-Hastings algorithm \citep{metropolis,hastings}. These inverse solutions are used to present reliable error estimates for all the model parameters in the form of Bayesian confidence sets. As a result, we present strong evidence for a new and previously unknown planetary companion orbiting HD 11506. We also demonstrate the importance of treating the magnitude of stellar RV noise, the RV jitter, as a free parameter in the model.

\section{The model and Bayesian model comparison}

When assuming the gravitational interactions between the planetary
companions to be negligible, the RV measurements with $k$ such companions can be modelled as \citep[e.g.][]{green}
\begin{eqnarray}\label{model}
  \dot{z}(t) & = & \sum_{j=1}^{k} K_{j} \big[ \cos ( \nu_{j}(t) + \omega_{j} ) + e_{j} \cos \omega_{j} \big] ,
\end{eqnarray}
where $\nu_{j}$ is the true anomaly, $K_{j}$ is the RV semi-amplitude, $e_{j}$ is the eccentricity, and $\omega_{j}$ is the longitude of pericentre. Index $j$ refers to the $j$th planetary companion. Hence, the RV signal of the $j$th companion is fully described using five parameters, $K_{j}$, $\omega_{j}$, $e_{j}$, mean anomaly $M_{0,j}$, and the orbital period $P_{j}$.

Following \citet{gregory05a,gregory07a,gregory07b}, we calculate the
Bayesian model probabilities for different statistical models. These
models include the first and second companion Keplerian signals and an
additive RV jitter (models M1 and M2, respectively). This jitter is
assumed to be Gaussian noise with a zero mean and its deviation,
$\sigma_{J}$, is another free parameter in the model. The jitter
cannot be expected to provide an accurate description of the RV noise
caused by the stellar surface. It also includes the possible
signatures of additional planetary companions if their signals cannot
be extracted from the measurements. Hence, this parameter represents
the upper limit to the true RV variations at the stellar surface. We also test a model without companions (M0) for comparison.

The instrument error, whose magnitude is usually known reasonably
well, is also included as a Gaussian random variable in our analysis.
Hence, for a model with $k$ planetary companions, the RV measurements $r_{i}$ at $t_{i}$ are described by
\begin{equation}\label{RV_model}
  r_{i} = \dot{z}_{k}(t_{i}) + \gamma + \epsilon_{I} + \epsilon_{J} ,
\end{equation}
where $\epsilon_{I} \sim N(0, \sigma_{I}^{2})$ is the instrument error, $\epsilon_{J} \sim N(0, \sigma_{J}^{2})$ describes the remaining uncertainty in the measurements, called the RV jitter, and parameter $\gamma$ is a reference velocity parameter.

When deciding whether a signal of a planetary companion has been detected or not, we adopt the Bayesian model selection criterion \citep{jeffreys}. This criterion states that a model with $k+1$ planetary companions should be used instead of one with $k\,$ companions if
\begin{equation}\label{best_model}
  P(\dot{z}_{k+1} | \mathbf{r}) \gg P(\dot{z}_{k} | \mathbf{r}) ,
\end{equation}
where $\mathbf{r}\,$ is a vector consisting of the RV measurements.
For more information about Bayesian model comparison, see \citet[e.g.][]{kass,gregory05a}.

After identifying the most appropriate model in the context of Eq.
(\ref{best_model}), we must find the Bayesian credibility sets that
can assess accurately the uncertainties in the orbital parameters and
the RV masses of the planetary companions. The Bayesian credibility
sets are robust uncertainty estimates because they show the
uncertainty of the model parameters given the measurements. A Bayesian credibility set $\mathcal{D}_{\delta}$ containing a fraction $\delta \in [0,1]$ of the probability density of parameter vector $\vec{u}$ can be defined to a subset of the parameter space $U$ that satisfies the criteria \citep[e.g.][]{kaipio}
\begin{equation}\label{Bayesian_confidence_set}
  \left\{ \begin{array}{l}
    \int_{\vec{u} \in \mathcal{D}_{\delta}} p(\vec{u} | m) d \vec{u} = \delta \\
    p(\vec{u} | m) | _{\vec{u} \in \partial \mathcal{D}_{\delta}} = c,
  \end{array} \right. 
\end{equation}
where $p(\vec{u} | m)$ is the probability density of the parameters, $c$ is a constant, $m$ is the measurements, and $\partial \mathcal{D}_{\delta}$ is the edge of the set $\mathcal{D}_{\delta}$. We use $\delta = 0.99$ throughout this article when discussing the parameter errors.

The question of model comparison is not only statistical but also
physical. Because the inverse solution is based on non-interacting
planets -- a simplification that is fairly adequate in most cases --
it might be physically impossible in reality. This is important to
understand because many extrasolar systems contain large
eccentricities that may cause close encounters. In these cases, a
detailed description of the dynamics is required. We drew 50 random
values of $\vec{u}$ from its posterior probability density and
integrated these directly to investigate the orbital evolution. The
integration was conducted using the Bulirsch-Stoer integrator
\citep{bulirsch}, which is famous for its high reliability even when
the dynamics include several close encounters. The stability analysis
that we used in this analysis is not exact but estimates reliably the
expected long-term behaviour from a relatively short ensemble or
orbit. In this simple analysis, we identify the variation in the
semi-major axis and the eccentricity from randomly selected initial
conditions to the end of integration. If the variations are not
significant the system is then assumed to be long-term stable (N.
Haghighipour, priv. comm.). The advantage this formulation is the
short integration time required when compared to more adequate, for
example, Lyapunov exponent based, integration methods.

The most important error source in this type of study is a phenomenon
known as stellar jitter. It is caused by a combination of convection,
rotation, and magnetic activity on the stellar surface \citep[see
e.g.,][and references therein]{wright05}. Although the role of stellar
activity as an error source is well known, the magnitude of this error
is almost always assumed to be constant, based on certain studies
\citep{wright05}. Here, we use a more conservative approach and
consider the magnitude of the jitter to be free parameter.

Before this work, four exoplanets were found using Bayesian approach
instead of a more traditional periodogram, the first of them being HD
73526 c \citep{gregory05a}. The candidate HD 208487 c was proposed by
\citet{gregory05b} and later confirmed by additional observations
\citep{butler, gregory07a}. For the system HD 11964, the Bayesian
analysis revealed evidence of three planets, despite only one being
previously known \citep{gregory07b}. None of these works include a
discussion of the dynamics, although it is clear that including
dynamics in the work enhance the quality of results. This point also
is interesting because if the dynamical analysis excludes a part of
the parameter space as physically impossible, this restriction can be
inserted into Bayesian model as additional \emph{a priori} constraint
that will, with the data, provide tighter confidence limits.

\section{Orbital solutions}

The star HD 11506 is a quiescent main sequence star of spectral class
G0 V. It is a relatively nearby star with a Hipparcos parallax
18.58~mas, which corresponds to a distance of 53.8~pc. It has $T_{eff}
= 6060$~K and [Fe/H] = 0.31 \citep{valenti} and its mass is estimated
to be 1.19~M$_{\odot}$ \citep{fischer}. We use this mass estimate
throughout the paper when calculating planetary masses.

The planetary companion HD 11506 b was first announced by
\citet{fischer}. They speculated an additional companion could be
present because the $\chi^{2}\,$ value of their single-companion model
fit was large (10.3). However, their $\chi^{2}$ value was calculated
by assuming a fixed jitter level. \citet{fischer} assumed that $\sigma_{J} = 2.0$ms$^{-1}$. Our solution for this parameter is consistent with this estimate (Table \ref{orbital_solutions}).

The Bayesian model probabilities for $k=1, 2$ are listed in Table
\ref{Bayesian_probabilities}. These probabilities imply that M2
provides tha most accurate description of the RV data of HD 11506.
These probabilities favour two planetary companions, implying that,
according to the measurements of \citet{fischer}, this star does host
two companions. It also verifies that the large $\chi^{2}$ value was
consistent with there being a signal of an additional companion, as
suspected by \citet{fischer}. The corresponding $\chi^{2}$ value of
our two-companion solution is 3.1. They also mentioned that a 170 day
period could be present in the data but were unable to verify the existence of a second companion. This period corresponds to our two-companion solution (Table \ref{orbital_solutions}).

When assuming the fixed jitter level adopted by \citet{fischer}, the
probability of the one-companion model was again found to be
considerably lower than in Table \ref{Bayesian_probabilities} (less
than $10^{-36}$). This result emphasizes the fact that the jitter
cannot be fixed but must be considered as a free parameter.
Furthermore, by fixing the jitter to some \emph{a priori} estimated
value, the probability densities of the orbital parameters become far
narrower, which underestimates the uncertainty in the solution.

\begin{table}
\center
\caption{Bayesian model probabilities of 1 and 2 planet models.\label{Bayesian_probabilities}}
\begin{tabular}{lc}
  \hline \hline
  Model & Probability \\
  & HD 11506 \\
  \hline
  M1 & $<10^{-6}$ \\
  M2 & 1 \\
  \hline
\end{tabular}
\end{table}

\begin{table}
\center
\caption{The RV two-planet solution of HD 11506. MAP estimates of the parameters and their $\mathcal{D}_{0.99}$ sets.\label{orbital_solutions}}
\begin{tabular}{lcc}
  \hline \hline
    Parameter & MAP & $\mathcal{D}_{0.99}$ \\
  \hline
    $P_{1}$ [yr] & 3.48 & [3.22, 4.01] \\
    $e_{1}$ & 0.22 & [0.10, 0.47] \\
    $K_{1}$ [ms$^{-1}$] & 57.4 & [49.7, 71.2] \\
    $\omega_{1}$ [rad] & 4.5 & [3.8, 5.1] \\
    $M_{1}$ [rad] & 4.7 & [2.5, 0.5] \\
    $m_{p,1} \sin i_{1}$ [M$_{Jup}$] & 3.44 & [2.97, 4.34] \\
    $a_{1}$ [AU] & 2.43 & [2.31, 2.67] \\
  \hline
    $P_{2}$ [yr] & 0.467 & [0.450, 0.476] \\
    $e_{2}$ & 0.42 & [0, 0.62] \\
    $K_{2}$ [ms$^{-1}$] & 25.5 & [11.2, 35.3] \\ 
    $\omega_{2}$ [rad] & 4.1 & [2.3, 5.1] \\
    $M_{2}$ [rad] & 5.5 & [0, 2$\pi$] \\
    $m_{p,2} \sin i_{2}$ [M$_{Jup}$] & 0.82 & [0.32, 1.13] \\
    $a_{2}$ [AU] & 0.639 & [0.622, 0.646] \\
  \hline
    $\gamma$ [ms$^{-1}$] & 6 & [-4, 15] \\
    $\sigma_{J}$ [ms$^{-1}$] & 3.5 & [0.6, 8.8] \\
  \hline
\end{tabular}
\end{table}

\begin{figure}
\begin{center}
\includegraphics[angle=270, width=8.0cm,
totalheight=6.0cm]{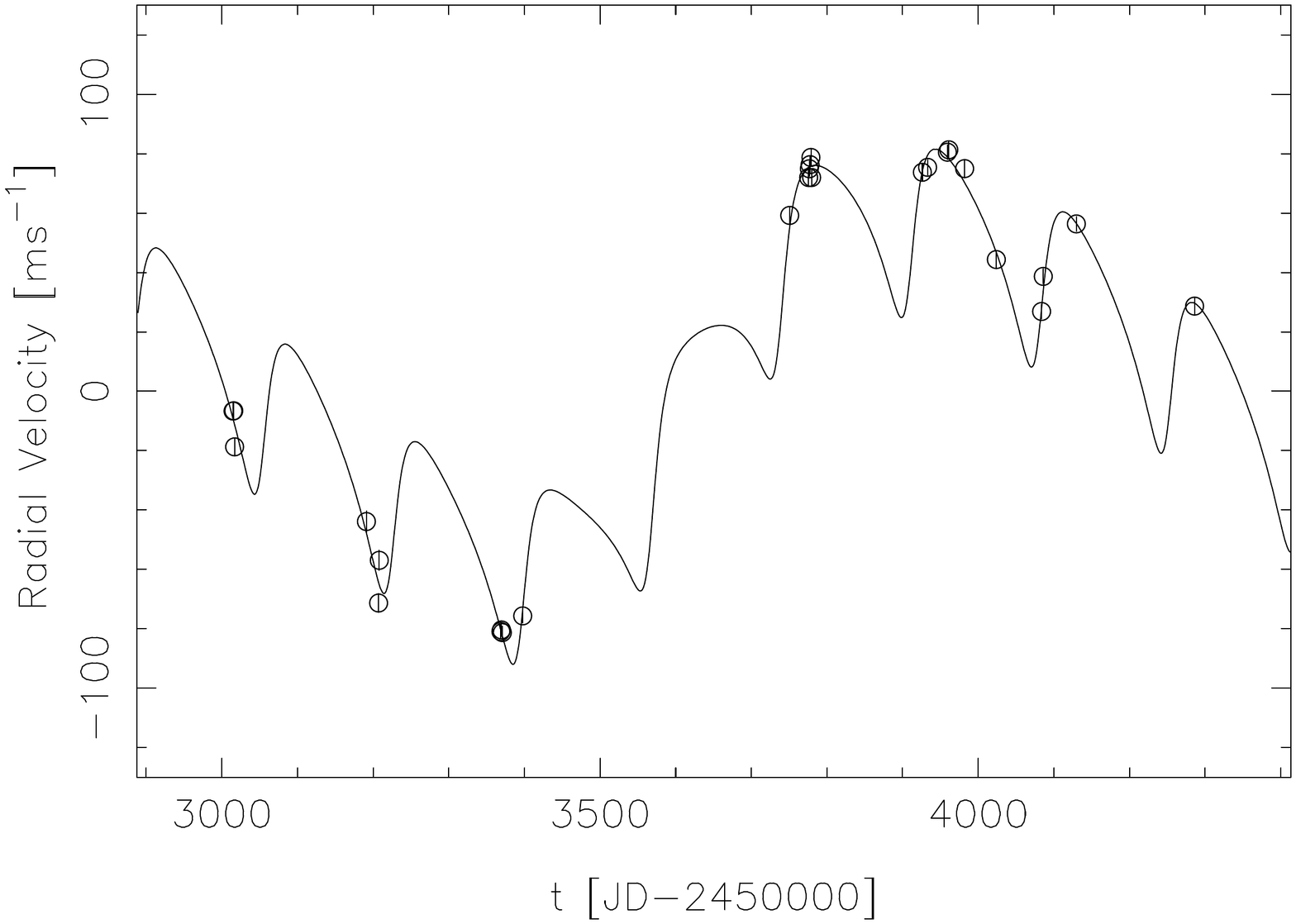}

\includegraphics[angle=270, width=8.0cm,
totalheight=6.0cm]{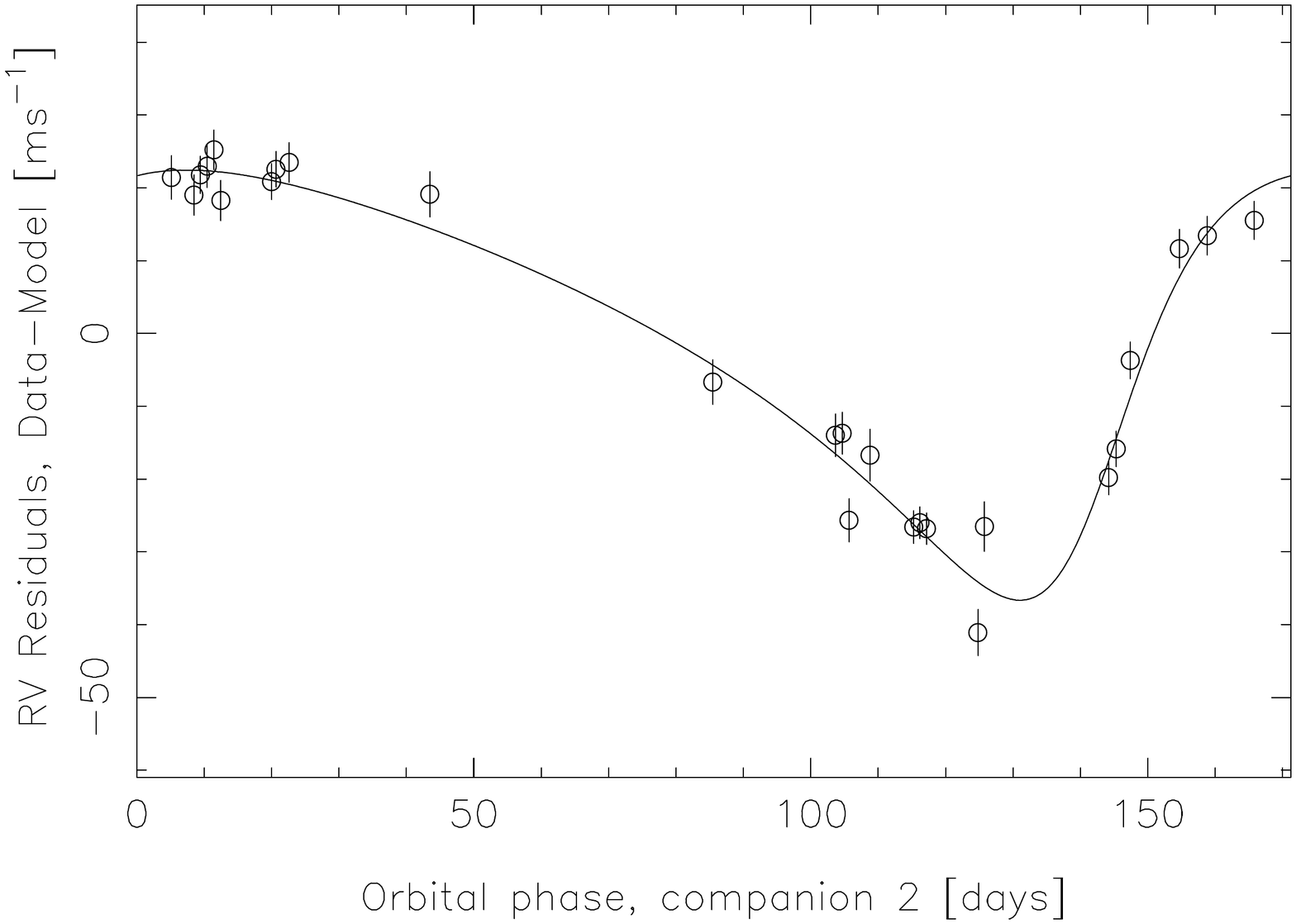}
\end{center}
\caption{Radial velocity measurements of HD11506 (Fischer et al. 2008). Two companion MAP orbital solution (top), and signal of companion HD 11506 c (bottom).} \label{orbit_HD11506}
\end{figure}

On Fig. \ref{orbit_HD11506} we show the maximum \emph{a posteriori}
(MAP) orbital solution of model M2 and the velocity curve of HD 11506
c after the signal of b companion has been subtracted. Interestingly,
our solution for the companion HD 11506 b differs from that of
\citet{fischer}. For instance, we found its RV mass to be
3.4M$_{Jup}$, whereas they reported a mass of 4.7M$_{Jup}$, which is
outside the margins of $\mathcal{D}_{0.99}$ set in Table
\ref{orbital_solutions}. The jitter parameter also appears to have a
higher value than estimated. This could be indicative of the
difficulties in estimating the jitter magnitude but the jitter may
also contain the signal of a third companion.

\begin{figure}
\begin{center}
\includegraphics[angle=270,
totalheight=6.0cm]{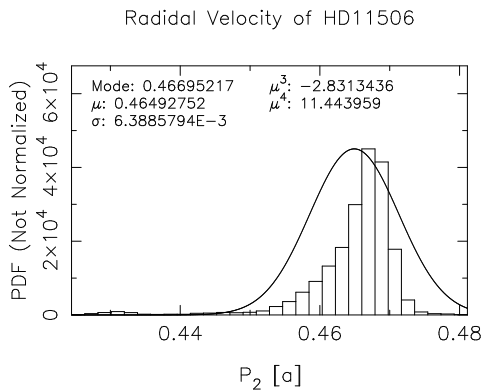}
\end{center}
\caption{The posterior probability density of parameter $P_{2}$ and its mode, mean, deviation, and two higher moments. The solid curve is a Gaussian curve with the mean and variance of the density.} \label{parameter_p2}
\end{figure}

We were unable to determine any strong correlations between the model
parameters, and all of the parameter densities, apart from $P_{2}$ (Fig. \ref{parameter_p2}), were reasonably close to Gaussian.

We subtracted the MAP one-planet solution from the RV measurements and
calculated the Scargle-Lomb periodogram \citep{scargle,lomb} for these
residuals. The highest peak corresponded to a period of 0.45~yr, which
is close to the period of the second companion (0.47~yr). However, the
FAP of this peak was as high as 0.54, which ensures that it was
impossible to detect the signal of this companion with periodogram. In
contrast, this solution was easily found using the MCMC method. In
agreement with the MAP solution, we were unable to find other probability maxima in the parameter space of the two-companion model.

\section{Orbital stability}

\begin{figure}
\begin{center}
\includegraphics[angle=0, totalheight=4.0cm]{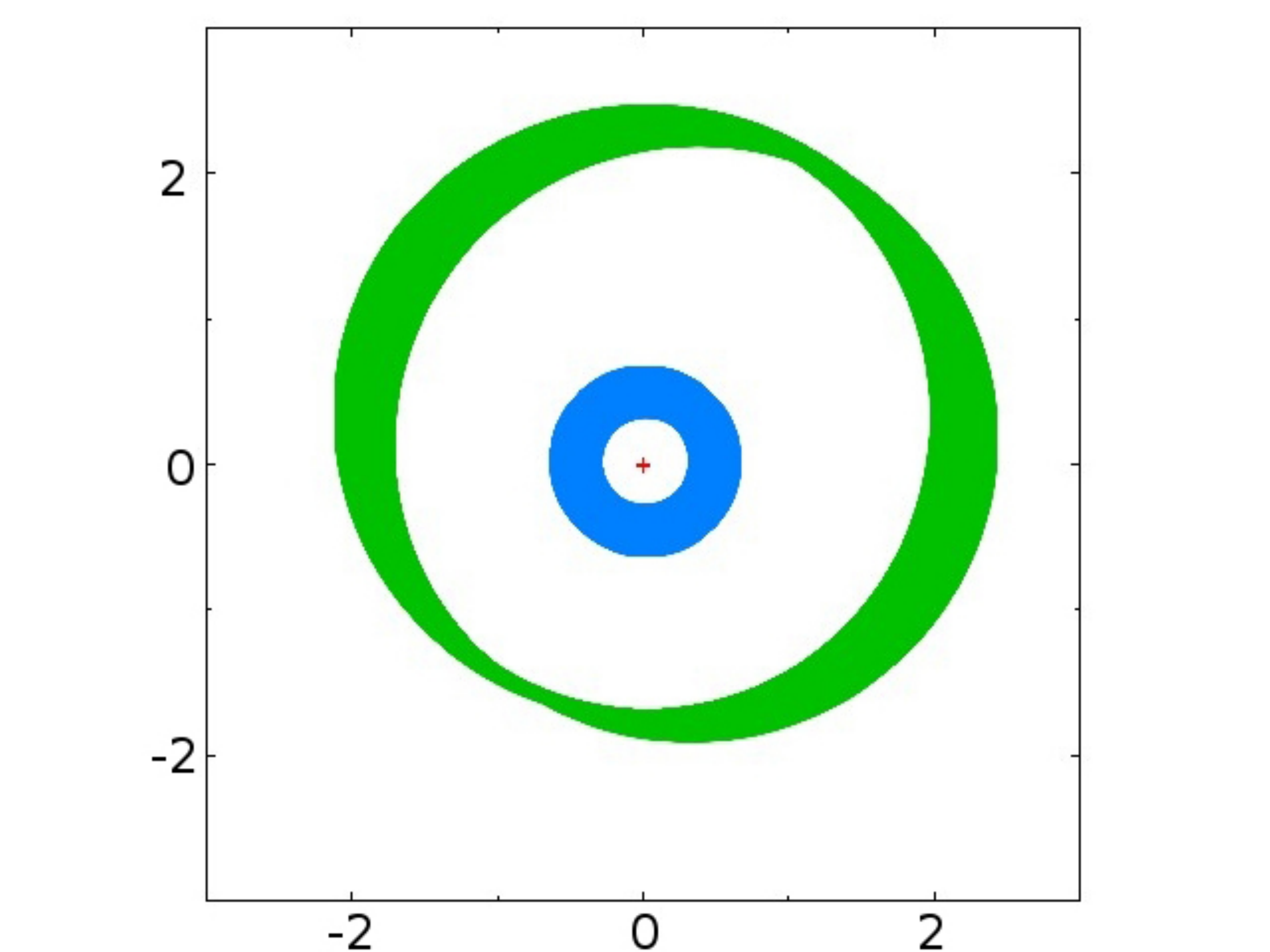}
\includegraphics[angle=0, totalheight=4.0cm]{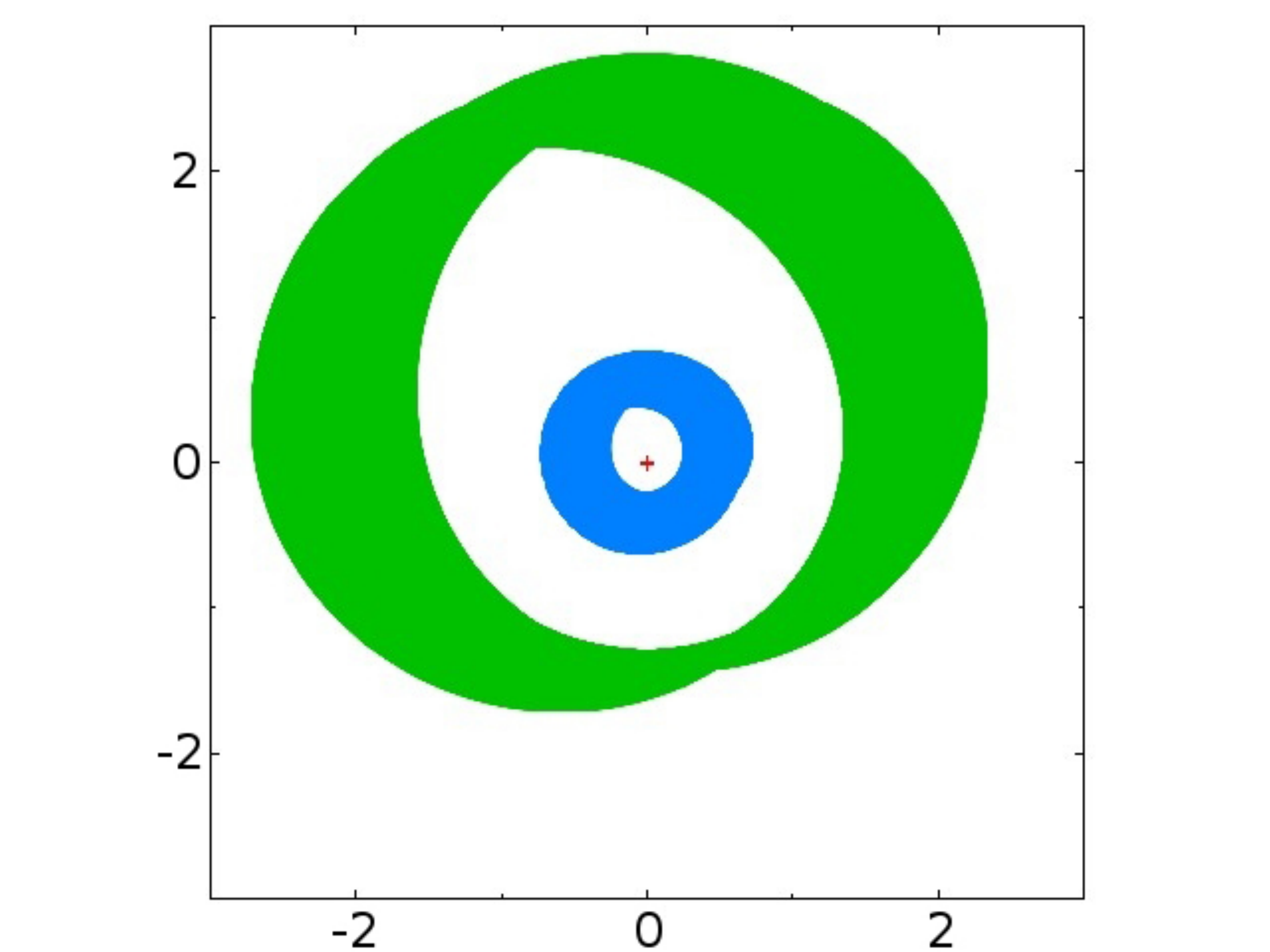}
\end{center}
\caption{Two random examples of two-planet solution for HD 11506.
Green and blue areas indicate orbits of HD 11506 b and c,
respectively. Red cross denotes the star. Both
figures present a 100~000 year part of orbital evolution, and the slow
precession of the apsid line is clearly visible. Unit of both axes is AU.} \label{simulations_HD11506}
\end{figure}

To present additional evidence of HD 11506
c, we selected 50 random combinations of parameters taken from the
parameter probability densities. These values were used to simulate
the dynamical behaviour of the planetary system by direct integration.
A random example of these simulations is shown in Fig.
\ref{simulations_HD11506} (top), where 100~000~yr excerpts are presented for
two randomly selected parameter combinations. The orbital ellipses
precess slowly but both the semimajor axis and the eccentricities remain
almost constant during the evolution. This feature does not prove but
suggests strongly that the two planetary companions orbiting HD 11506
provide a physically stable system.

We tested the stability further by selecting 50 values of the
parameter vector from the region of parameter space most likely
prone to instability, i.e., where the quantity $Q =
a_{1}(1-e_{1})/(a_{2}(1+e_{2}))$ has its smallest value. We drew the values
from a probability density, whose maximum is at $\min Q$ and that
decreases linearly to zero at $\max Q$. This results in a sample from
the posterior density that has values mainly close to $\min Q$. This
test was designed to obtain further information about the orbital
parameters by excluding parameter values that excluded an unstable
system. However, all 50 parameter values resulted in bound orbits
even though they precessed strongly. Therefore, it was impossible to
extract additional information about the orbital parameters
from these simulations. See random example in Fig.
\ref{simulations_HD11506} (bottom).

\section{Conclusions}

We have presented complete Bayesian reanalysis of radial velocities of
HD 11506, and identified the orbital parameters of a previously unknown exoplanet candidate.

These analyses demonstrated the importance of taking stellar jitter,
the most important error source in RV measurements, into account as a
free parameter in the analysis. As an unknown noise parameter, jitter
cannot be predetermined to some estimated value because of its strong
effect on the Bayesian model probabilities of models with different
numbers of companions. If overestimated, it may prevent the detection
of a additional companions. Its uncertainty must also be taken into account when
calculating the error bars for the orbital parameters to prevent the
underestimation of their errors.

In the case of HD 11506 the two-planet model is by far more probable
for the given data \citep{fischer} than the original one-planet
solution, a result that remained unchanged regardless of whether we
considered the jitter as fixed or free. The fit of the two-planet
model is similar to any known two-planet signal and a dynamical
analysis has demonstrated the solution to be physically possible. We
therefore claim that the RV data, which was presented by \citet{fischer}, contains the signals of two planetary companions.

Another interesting feature of the full inverse solution of the
two-companion model is that the RV mass of HD 11506 b differs
significantly from that obtained by \citet{fischer}. This difference
implies that the Bayesian model selection criterion should be used
when assessing the number of planetary signals in RV data. Without
accurate knowledge on the best-fit model, the orbital solution can be
biased and the resulting statistical conclusions about the properties of extrasolar planetary systems can be misleading.

More observations are required to tighten further the constraints on the parameter space of HD 11506 system.

\section*{Acknowledgements}
S.K. has obtained funding from Jenny and Antti Wihuri Foundation. The
authors would like to thank the referee for valuable
suggestions and comments and to acknowledge Proffan Kellari for
inspiring environment during the group meetings.


\end{document}